\documentclass[]{aa}
\usepackage{psfig}
\begin{document}

\title{A variable star survey of the open cluster M37}

\author{L.L. Kiss\inst{1,4} \and Gy. Szab\'o\inst{1} \and 
K. Szil\'adi\inst{1,4} \and G. F\H{u}r\'esz\inst{1} \and
K. S\'arneczky\inst{2,4} \and B. Cs\'ak\inst{1,3}}

\institute{Department of Experimental Physics and Astronomical
Observatory, University of Szeged, D\'om t\'er 9., H-6720 Hungary \and
Department of Physical Geography, ELTE University, H-1088 Budapest,
Ludovika t\'er 2., Hungary \and
Department of Optics \& Quantum Electronics, University of Szeged,
POB 406, H-6701 Szeged, Hungary \and
Guest Observer at Konkoly Observatory}

\titlerunning{Variable stars in M37}
\authorrunning{Kiss et al.}
\offprints{L.L. Kiss \\
\email{l.kiss@physx.u-szeged.hu}}
\date{}

\abstract{
A CCD photometric study of the dense galactic open cluster M37 is presented
and discussed. The majority of the analysed data are time-series
measurements obtained through an R$_C$ filter. The observations were carried
out on seven nights between December 1999 and February 2000, and have led
to the discovery of 7 new variable stars in the field. Three of them have
been unambiguously identified as W UMa-type eclipsing binaries, while two
more are monoperiodic pulsating stars, most probably high-amplitude $\delta$
Scuti-type variables. The remaining two stars seem to be long-period
eclipsing binaries without firm period determination. Johnson B and V
frames have been used to construct a new colour-magnitude (CM) diagram of
the cluster, and to find the locations of the new variable stars. The
pulsating variables are most likely background objects. The CM diagram is
fitted with recent isochrones yielding the main parameters of the
cluster.
}

\maketitle

\section{Introduction}

Variable stars in open clusters provide an important tool to test
theoretical predictions concerning stellar structure and evolution, because
the same distance, age, initial chemical abundance and interstellar
reddening can be assumed for all the stars in the same cluster. Moreover,
studying cluster variables is one of the most effective and productive uses
of small and moderate size telescopes. Simultaneous photometry of hundreds
of stars enables very accurate light curves to be obtained. In the recent
years, substantial development has been made in this area, owing to the
importance of, for instance, studying stellar pulsation (Frandsen \&
Arentoft 1998) or evolution of close binaries (Ka\l u\.zny \& Rucinski 1993)
in open clusters. Evolutionary effects make different types of variable
stars appear in specific range of cluster ages, therefore, it is highly
desirable to make observations of as many clusters with different ages as
possible. The function of this paper is to contribute to this issue with
new observations of the intermediate-age open cluster M37.

M37 (=NGC 2099=C0549+325, $\alpha_{2000}=05^h52\fm3, \delta_{2000}
=+32^\circ33^\prime$) is a very rich open cluster located in the low
galactic latitude (b=+3\fdg09) region in the constellation Auriga. Although
it is among the brightest open clusters in the northern sky, the
observational material in the literature is quite scanty. The last extensive
photometry is that of West (1967) who presented photographic UBV
measurements for 930 stars in the cluster. The main sequence is rather
broad and the available proper motions (Jefferys 1962, Upgren 1966) do not
enable a clear distinction between the member and non-member stars. Robin
(1982) published electronographic plate measurements for selected stars in
the cluster in the UBV bands for calibration purposes. The fifth catalogue
of Lyng\aa\ (1987) lists the following cluster parameters: m$-$M=11\fm63,
E(B$-$V)=0\fm31 and $log~{\rm t}$=8.30. The most recent update of these
parameters was given by Mermilliod et al. (1996) who studied the red giant
content of this cluster and found m$-$M=11\fm50, E(B$-$V)=0\fm29 and
$log~{\rm t}$=8.65. In contrast to the above, the WEBDA catalogue ({\tt
http://obswww.unige.ch/webda}) lists m$-$M=11\fm67, E(B$-$V)=0\fm302,
$log~{\rm t}$=8.540, however, the differences of the quoted values do not
exceed the expected uncertainties. To our knowledge, there has been no
variable star survey in the literature despite the relative proximity of the
cluster. This neglect might be caused by the strong concentration of stars
that makes it difficult to separate individual stars.

\section{Observations and data analysis}

CCD BVR photometric observations were carried out on 7 nights between
December 27, 1999 and February 1, 2000 at the Piszk\'estet\H{o} Station of
the Konkoly Observatory (Hungary). The instrument used was the 60/90/180 cm
Schmidt telescope, equipped with a Photometrics AT-200 CCD (1536$\times$1024
pixels, KAF-1600 chip with UV-coating). The angular resolution was
1\farcs1/pixel, yielding a field of view of 29$^\prime \times 18^\prime$.
Time-series observations were made through an R filter (60 and 120 s
exposures), while three B and V filtered frames were also taken on December
29, 2000 in order to obtain colour information on the detected stars (20
s, 60 s and 300 s exposures). The choice of using an R filter in
the time-series observations was due to an unfortunate technical
difficulty, that made other filters unavailable on other nights. The full
observing log is given in Table\ 1.

%Table 1
\begin{table}
\caption{The journal of observations}
\begin{center}
\begin{tabular} {lcccl}
\hline
Date       & B & V & R & length\\
           & frames   & frames   & frames   & (min)\\
\hline
1999 Dec. 27/28  & -- & -- & 120 & 190\\
1999 Dec. 29/30  & 3  & 3  & 30  & 100\\
1999 Dec. 30/31  & -- & -- & 187 & 280\\
2000 Jan. 28/29  & -- & -- & 145 & 340\\
2000 Jan. 30/31  & -- & -- & 43 & 90\\
2000 Jan. 31/Feb. 01 & -- & -- & 58 & 150\\
2000 Feb. 01/02 & -- & -- & 75 & 170\\
\hline
{\bf Total:}    & 3  & 3  & 658 & 1320\\
\hline
\end{tabular}
\end{center}
\end{table}

The basic image reduction was carried out using the standard tasks in IRAF.
The flat field corrections made use of sky-flat images taken during the
evening twilight. We performed psf-photometry with the {\it daophot}
package of IRAF using the Moffat point-spread function, which turned out to
be well suited for fitting the slightly distorted stellar profiles (caused
by the optical layout of the Schmidt system). The magnitudes were
calculated relative to the ensemble mean of a selected set of stars, with
an arbitrarily chosen zero-point. Typically, we detected about 3000 stars
on the R frames. The co-ordinate system of each frame was corrected for
two-dimensional shift, rotation and optical distortions, relative to a
master frame. The master frame is the one which contained the largest
number of stars, and is likely to contain every stars detected in the
series. Although there is still some small chance of loosing a few stars
that were undetected on the given starting image, the overwhelming majority
of the stars are properly identified. We could identify 2530 stars on at
least 20 frames. The last step in data filtering was to reject those stars
that were closer than 30 pixels to the image edges. The finally adopted data
set contains 2323 individual light curves. We note, that R data on Dec.
29/30, 1999 (30 points) were excluded from the photometric analysis because
of their larger noise level.

%Fig. 1.
\begin{figure}
\begin{center}
\leavevmode
\psfig{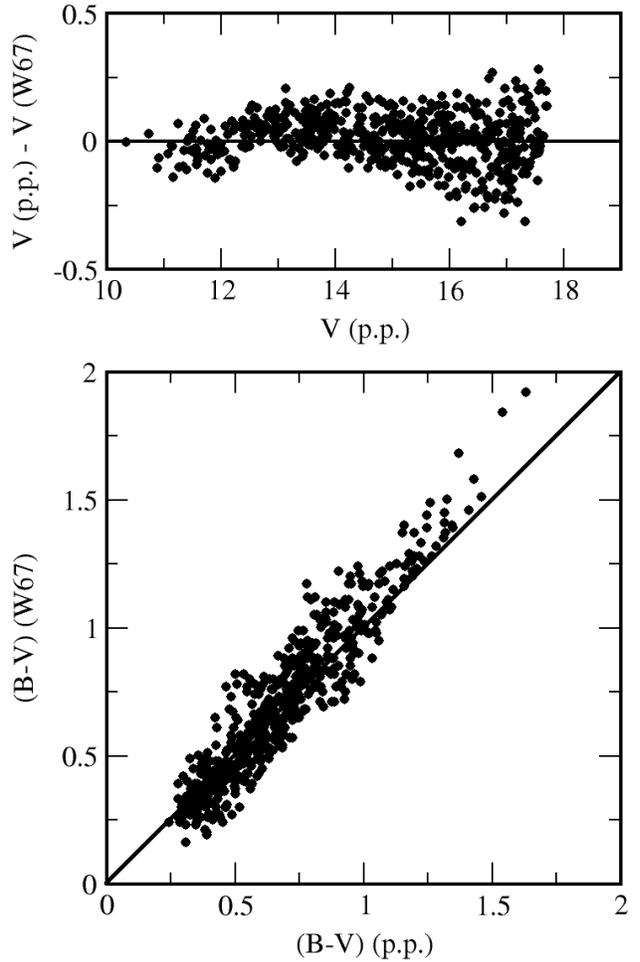}
\caption{A comparison of the V magnitudes and B$-$V colours obtained in this
study with data by West (1967). The diagrams are based on 580 common stars.
The solid lines denote equality.}
\end{center}
\label{f1}
\end{figure}

The weather conditions were not photometric when the three-colour
observations were obtained, therefore, we could not perform standard all-sky
photometry in order to standardise the observed colours and magnitudes.
Thus, we have tied our data to the photoelectric UBV photometric secondary
standards in the cluster taken from Table\ 2 in West (1967). The
colour-dependent transformation coefficients $\epsilon$ and $\mu$ ($\Delta
V=\Delta v + \epsilon \Delta (B-V)$, $\Delta (B-V)=\mu \Delta (b-v)$) were
determined by observing the standard sequence of Schild (1983) in the open
cluster M67 ($\epsilon$=0.10$\pm$0.03, $\mu$= 0.83$\pm$0.03).
This $\mu$ value indicates filters or a detector quite far from
the standard, fortunately, the standard star observations gave
well-defined linear transformation equations with no indication of
second-order terms. We have
compared the resulting V and B$-$V data with those of West (1967) and can
unambiguously identify 580 stars common to both sets which are plotted in
Fig.\ 1. There are some slight systematic differences between the V data
which are characterized by the mean deviation of V(p.p.)--V(W67) 
and the rms around that mean in certain magnitude ranges. 
The results are: for V=10--12 mag the mean deviation is $-0\fm032$ with
an rms 0\fm043; for V=12--14 mag +0\fm044, 0\fm049; for V=14--16 mag 
+0\fm015, 0\fm067; for V=16--18 mag $-0\fm013$, 0\fm098. 
The colour data are also affected by these systematic differences.
The largest deviations occur for the reddest stars, our values are slightly
bluer than those of West (1967). This inconsistency is due to the fact that
red stars beyond B$-$V$\approx$1\fm50 were essentially extrapolated with the
standard transformations. The two reddest photoelectric standards in Table\
2 of West (1967) have B$-$V 1\fm626 and 1\fm559 and such red stars are very
likely variable stars (Jorissen et al. 1997). In summary, the
uncertainty of standard V and B$-$V data changes from 0\fm03 to 0\fm10,
depending on the colour and apparent brightness of the stars.
Since we could not standardise the R-band data, their actual values
contain the arbitrary zero-point.

%Fig. 2.
\begin{figure}
\begin{center}
\leavevmode
\psfig{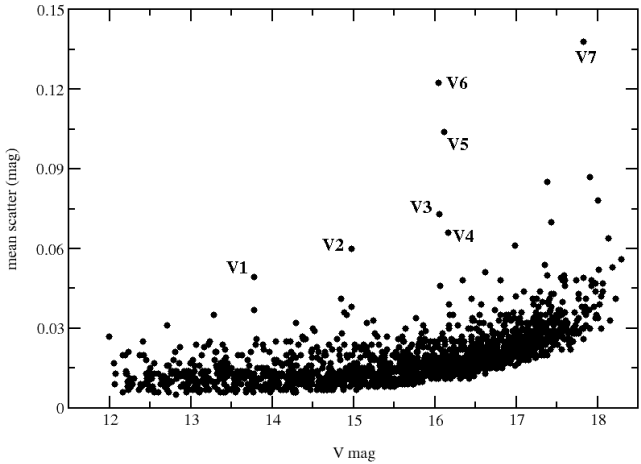}
\caption{The scatter of individual stars as a function of V magnitude. Note
the absence of the brightest stars (between V=11\fm0...12\fm0) in the
cluster field, which were saturated in the time series data observations.}
\end{center}
\label{f2}
\end{figure}

%Fig. 3.
\begin{figure*}
\begin{center}
\leavevmode
\psfig{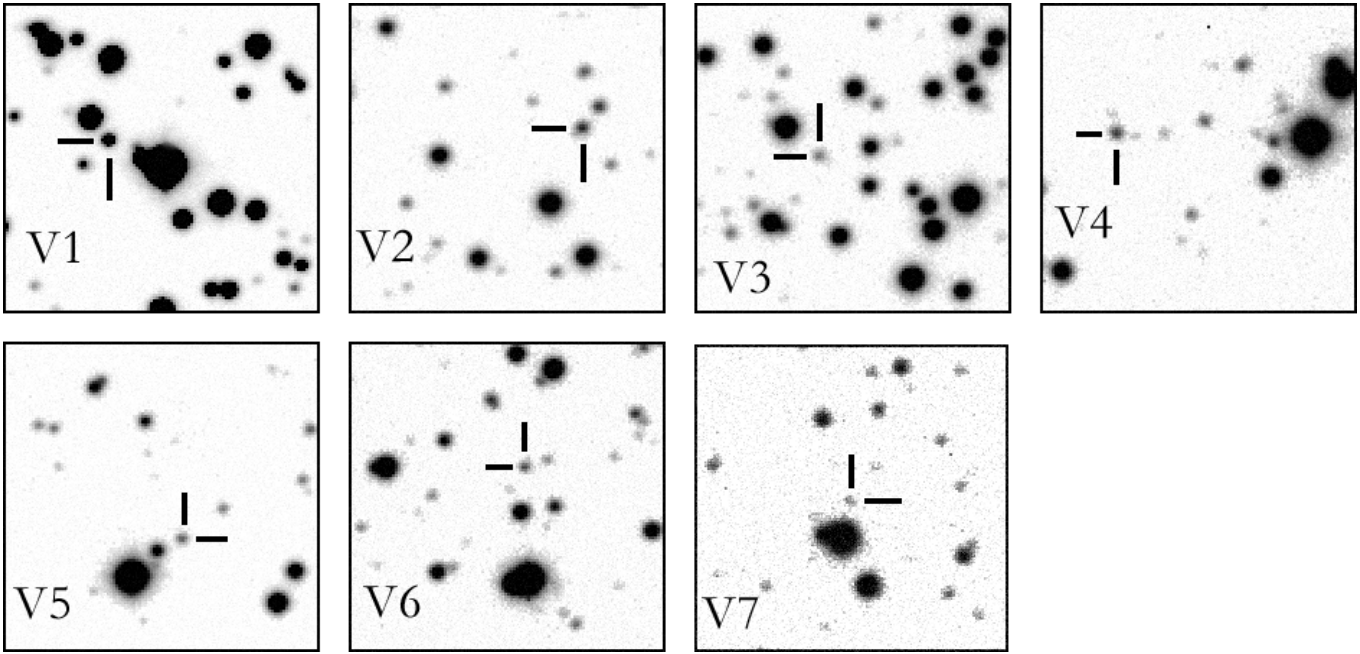}
\caption{Finding charts for the newly discovered variable stars. Each
frames are $2\times2^\prime$ wide, north is up, east is to the left.}
\end{center}
\label{f3}
\end{figure*}

The search for variable stars was performed by checking the light curve
statistics and by visual inspection of the suspected variable time-series
data. We
have calculated the scatter of individual light curves for every star (in
sense of $\langle m_{\rm obs} - \langle m_{\rm obs} \rangle \rangle$) and it
is plotted as a function of V magnitude in Fig.\ 2. The identified new
variables are also labelled. The main ridge helped us to characterise the
accuracy of the brightness measurements, which changed between
0\fm01-0\fm06 over a range of $\sim$6 magnitudes. As can be seen in
Fig.\ 2, besides the strongly deviant points, there is a group of slightly
deviant points over the main concentration. We have checked all of them and
found that those stars are in the most crowded fields in the cluster, where
the applied psf-photometry may suffer from larger photometric errors. That
is why all of the data with suggested larger scatter was inspected visually.
Seven stars remained as unambiguously variable stars and these are
identified in the $2\times2^\prime$ finding charts in Fig.\ 3.

The light curves of new variable stars were further analysed to search for
possible periodicities. For this, we performed conventional Fourier
analysis of the data with Period98 (Sperl 1998). However, in case of
eclipsing binaries, the phase dispersion minimisation (PDM - Stellingwerf
1978) is more suitable for inferring periodicities. Thus, PDM spectra
enabled an independent check of the results. Original data are available
electronically at CDS-Strasbourg.

\section{Results}

The analysis consisted of the following steps: {\it i)} construction of the
colour-magnitude (CM) diagram and physical parameter estimation of the
cluster; {\it ii)} variable star classification based on the light curve
shapes; {\it iii)} period determination for stars with well-sampled light
curves; {\it iv)} membership investigation (considering the location of new
variables on the CM diagram and using various period-colour-luminosity
relations that are valid either for W UMa or $\delta$ Scuti stars).

\subsection{The colour-magnitude diagram}

%Fig. 4.
\begin{figure*}
\begin{center}
\leavevmode
\psfig{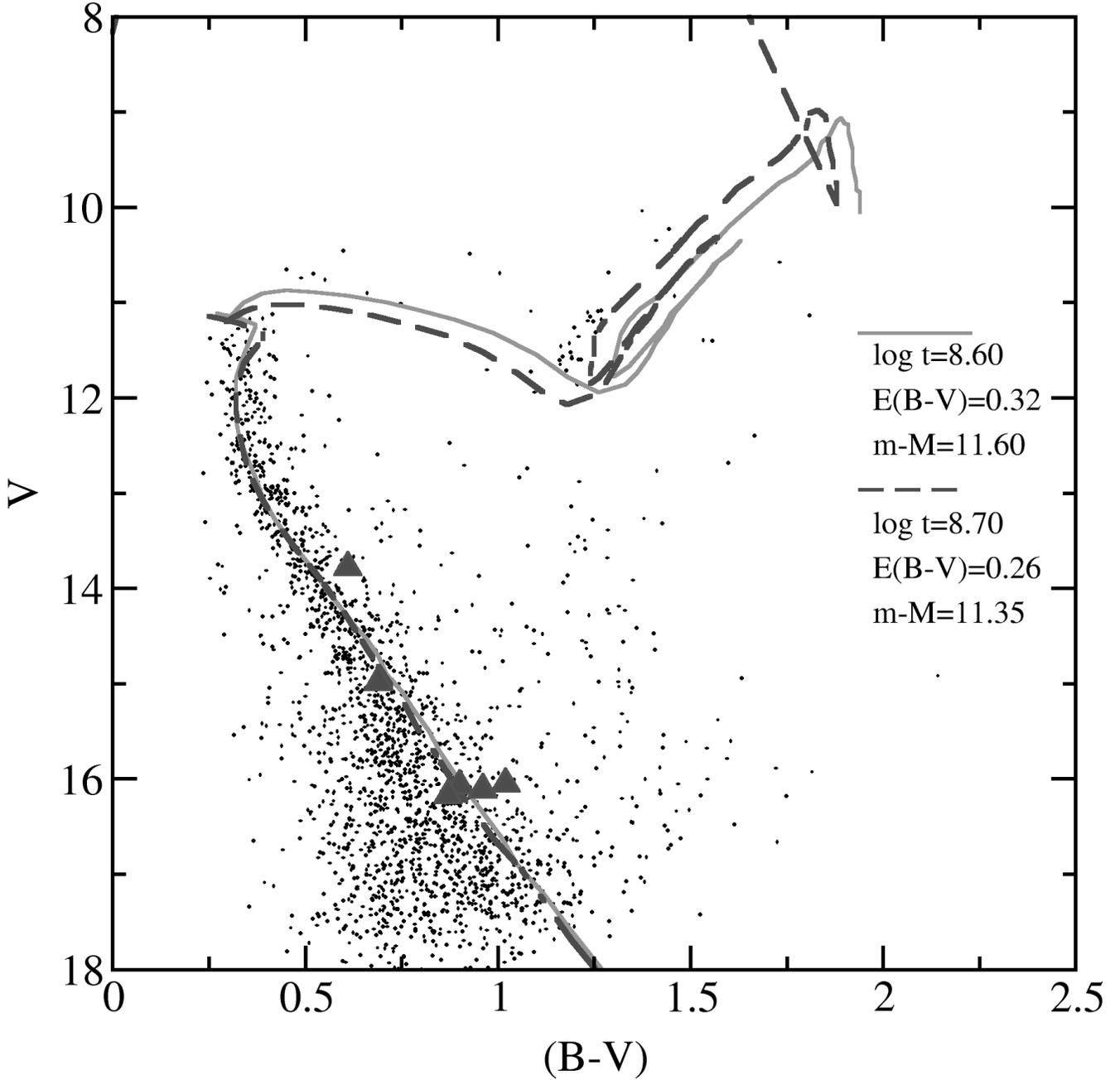}
\caption{The colour-magnitude diagram of M37 with two isochrones (solar
composition) taken from Bertelli et al. (1994). The morphology is well
reproduced, although the red giant models are slightly too red. The
positions of new variable stars are marked with the up triangles. The
brightest one is V1 (type EA:), the next one is V2 (EA:), while the four
variables at V$\approx$ 16 mag are V4 (EW), V3 (EW), V5 (DSCT/RRc) and V6
(DSCT) starting from the bluest one. V7 is not presented in this diagram
because it was not detected on the B frames.}
\end{center}
\label{f4}
\end{figure*}

The physical parameters of the cluster have been estimated by a  isochrone
fitting of the CM diagram. Since we could not determine the interstellar
reddening independently this parameter has been also included in the
fitting procedure. The isochrones were taken from Bertelli et al. (1994) and
the initial parameters were adopted from Mermilliod et al. (1996). The
result is shown in Fig.\ 4, where the CM diagram is plotted with the two
closest isochrones z=0.02). Note the strong field star contamination, which
is caused by the position of the cluster in the galactic plane. The overall
morphology is well reproduced, except that the models are too red for the
red giant stars. Note that this diagram is essentially the same as Fig.\ 11
in Mermilliod et al. (1996), the only difference is that our graph is based
on a much larger data set which extends to $\sim$2 mag below the faint
limit in Mermilliod et al. (1996). Therefore, the resulting parameters are
very similar to those of Mermilliod et al. (1996): the colour excess is
E(B$-$V)=0\fm29$\pm$0\fm03, the distance modulus is
m$-$M=11\fm48$\pm$0\fm13 and ${\rm log}~t$=8.65. These values were adopted
in this paper.

\subsection{The new variable stars}

%Fig. 5.
\begin{figure*}
\begin{center}
\leavevmode
\psfig{figure=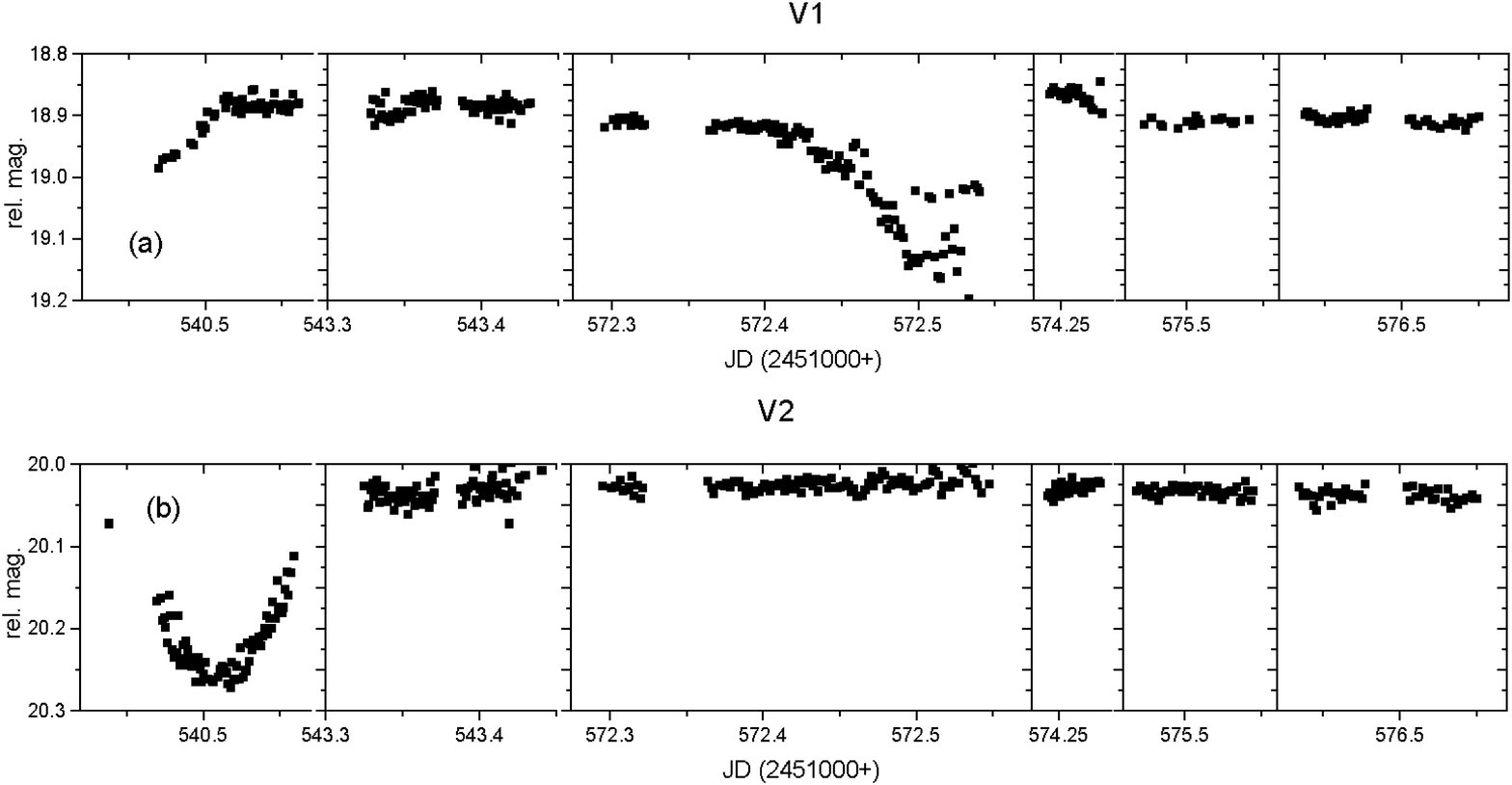,width=\textwidth}
\caption{The light curves for two new suspected Algol-type
eclipsing binaries.}
\end{center}
\label{f5}
\end{figure*}

In the following section we describe the new variables and their light
variations. Eclipsing binaries and pulsating variables are discussed
separately. Thanks to the moderately good phase coverage, five of the seven
new variables have unambiguous periods and classification. The remaining two
stars have characteristic light variation of long-period detached eclipsing
binaries, i.e. long constancy interrupted by short fadings. We have given
the stars the V1...V7 designations, where the numbers increase with the
average magnitude. Their basic data are summarised in Table\ 2.

%Fig. 6.
\begin{figure*}
\begin{center}
\leavevmode
\psfig{figure=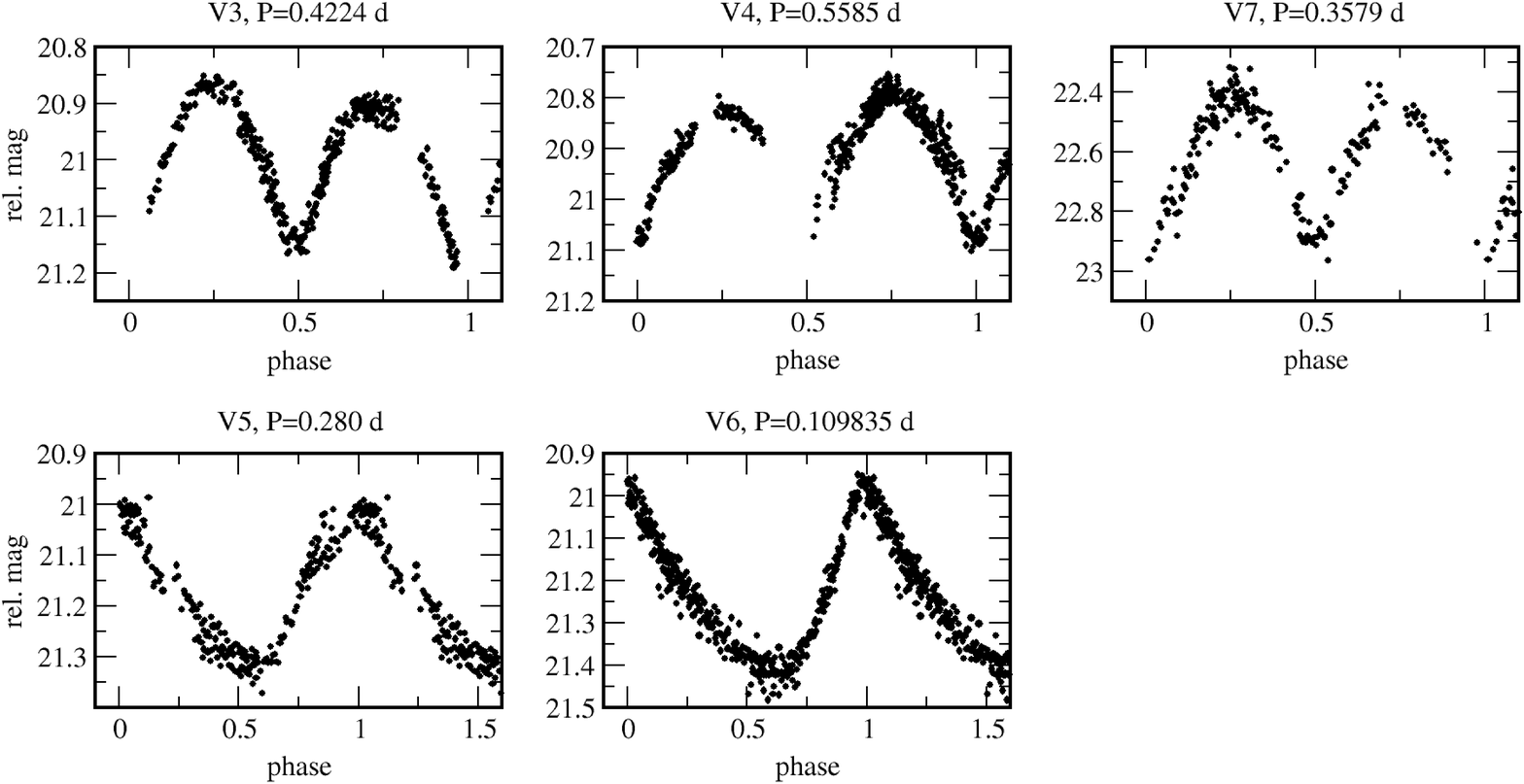,width=\textwidth}
\caption{The phase diagrams for V3..V7.}
\end{center}
\label{f6}
\end{figure*}

\subsubsection{Eclipsing binaries}

\noindent V1: There was only one fading of 0\fm24 on JD 2451572.
Unfortunately, we did not observe the subsequent brightening, thus no
epoch of minimum could be determined. The first night of observations
revealed an ambiguous brightening, which may be the ascending branch
of a primary minimum. The light curves are plotted in panel (a) of
Fig.\ 5. The larger scatter in subpanel a3 was caused by the bad
seeing which increased the light contamination from the close bright
neighbours (see its finding chart in Fig.\ 3).

\bigskip

\noindent V2: We observed a 0\fm23 deep minimum on JD 2451540 which
occured at HJD(min)=2451540.518. The other observations showed
no significant variation, thus no period could be determined.
The light curves are shown in panel (b) of Fig.\ 5.

\bigskip

\noindent V3: The amplitude is 0\fm32, while the inferred period is
0.4224$\pm$0.0001 days. The light curve shape is very characteristic for
a contact binary. The phase diagram of the whole light curve is shown in
Fig.\ 6. We note the slightly different maxima that might be a hint of a
spotted stellar surface.

\bigskip

\noindent V4: The light curve is typical of an EW-type star with an
amplitude of 0\fm33. There are interesting humps on the light curve
suggesting again the possibility of photometric surface inhomogeneities. The
inferred period is 0.5585$\pm$0.0001 days. The phase diagram is presented in
Fig.\ 6.

\bigskip

\noindent V7: This is the faintest variable star detected by us in the
cluster, and that is why its light curve contains only 185 points (it has
been lost from many frames by the star detection algorithm). The observed
amplitude is 0\fm55, while the period is about 0.3579$\pm$0.0001 days. The
phase diagram shows a typical W UMa-type variation (Fig.\ 6).

\subsubsection{Pulsating variables}

\noindent V5: The short-period asymmetric light curve with an amplitude of
0\fm32 is typical of pulsating stars. The period is 0.2800$\pm$0.0005 days.
Since it is located very close to the edge of the field, only 5 nights of
data were analysed and their phase diagram is shown in Fig.\ 6.

\bigskip

\noindent V6: The light variation is that of a monoperiodic high-amplitude
$\delta$ Scuti-star (P=0.109835$\pm$0.000005 days, A=0\fm45). The phase
diagram (Fig.\ 6) also suggests a remarkably repetitive light change.

We have also studied the possible multiperiodic nature of V5 and V6 by a
standard Fourier analysis of their light curves. After prewhitening with
the main frequencies (and their harmonics), the residual frequency spectra
did not show any significant peak. However, because of the relatively short
span of observations we cannot firmly conclude the strict monoperiodicity of
these variable stars.

%Table 2
\begin{table*}
\caption{The basic data of new variables}
\begin{center}
\begin{tabular} {lllllllll}
\hline
Star & $\alpha_{2000}$  & $\delta_{2000}$ & $\langle V \rangle$ &
$\langle B-V \rangle$ & P (d) & A (mag) & Epoch(s) & Type\\
\hline
V1 & 05$^h$52$^m$20\fs42 & +32$^\circ$33$^\prime$19\farcs5 & 13\fm78 &
0\fm61 & -- & -- & -- & EA:\\
V2 & 05$^h$52$^m$16\fs60 & +32$^\circ$28$^\prime$14\farcs9 & 14\fm98 &
0\fm69 & -- & -- & 2451540.518 & EA: \\
V3 &05$^h$52$^m$33\fs03 & +32$^\circ$32$^\prime$41\farcs7 & 16\fm07 &
0\fm90 & 0.4224(1) & 0.31 & 2451575.5083 (II) & EW \\
V4 & 05$^h$52$^m$53\fs26 & +32$^\circ$33$^\prime$01\farcs2 & 16\fm17 &
0\fm87 & 0.5585(1) & 0.33 & 2451576.456: (I) & EW \\
V5 & 05$^h$53$^m$00\fs63 & +32$^\circ$24$^\prime$50\farcs8 & 16\fm11 &
0\fm96 & 0.2800(5) & 0.32 & 2451576.500: & DSCT \\
V6 & 05$^h$51$^m$50\fs55 & +32$^\circ$32$^\prime$34\farcs4 & 16\fm05 &
1\fm02 & 0.109835(5) & 0.45 & 2451540.5367 & DSCT \\
%  & & & & &  & & 2451572.3908 & \\
%  & & & & &  & & 2451572.4997 & \\
%  & & & & &  & & 2451574.2578 & \\
%  & & & & &  & & 2451576.4541 & \\
V7 & 05$^h$52$^m$39\fs32 & +32$^\circ$36$^\prime$30\farcs9 & 17\fm83 &
-- & 0.3579(1) & 0.55 & 2451574.262: (I) & EW \\
\hline
\end{tabular}
\end{center}
\end{table*}

\subsection{Cluster membership of the variable stars}

We have marked the locations of the variable stars on the CM diagram in
Fig.\ 4 and must stress that these positions, based on a single-epoch
observation, are uncertain to $\sim\pm$0\fm2, as both the B$-$V colour and V
magnitude change over the time. This effect is larger for the pulsating
variables, which usually have more variable colour indices than eclipsing
stars. The locations of the eclipsing binaries are consistent with the stars
being members of the cluster, as they scatter around the main sequence. This
can be further studied for V3 and V4 with the period-colour-luminosity
relation published by Rucinski \& Duerbeck (1997). Their Eq. (5a)($M_{\rm
V}=-4.44 {\rm log}~P+3.02
(B-V)_0+0.12$) predicts $M_{\rm V}$(V3)=3\fm62 and $M_{\rm V}$(V4)=3\fm00
which is about 1-1.5 mag brighter than the calculated absolute magnitudes
using the cluster's distance modulus. Note, that when calculating the
absolute magnitude, the unreddened colour should be estimated, e.g., with
help of the period-colour diagram of W UMa-type systems (Fig.\ 2 in Rucinski
\& Duerbeck 1997).
However, the above quoted difference does not exlude the membership of V3
and V4, because even the calibrating data set of Rucinski \& Duerbeck (1997)
contained stars with similarly large deviations.

Contrary to the eclipsing binaries, the pulsating variables are too red and
too faint to be cluster members. The $\delta$ Scuti instability strip of the
H-R diagram (Breger 2000, Rodr\'\i guez \& Breger 2001) is in a bluer and
much brighter region than that containing V5 and V6 (e.g. the apparent
brightness range of the instability strip of M37 is about 11\fm5--14\fm0).

Petersen \& Christensen-Dalsgaard (1999) discussed
the Hipparcos period-luminosity relation for high-amplitude
$\delta$ Scuti stars and concluded that quite good
absolute magnitudes ($\pm$0\fm1) can be inferred
from the resulting PL-relation. Their Eq. (4)
($M_{\rm V}=-3.75 {\rm log}~P-1.969$)
predicts $M_{\rm V}$(V5)=0\fm1 and $M_{\rm V}$(V6)=1\fm6.
Both values suggest the stars are much further away than M37.
Unfortunately, farther means redder and the corresponding
higher colour excesses cannot be properly estimated.

Furthermore, the period and light curve shape of V5 also suggests the
possibility of being a field RRc star instead of high-amplitude $\delta$
Scuti star. In that case its absolute magnitude should be about 0\fm5 (${\rm
log}~L\approx1.7$, e.g. Koll\'ath et al. 2000), which yields to a distance
modulus m$-$M(V5)=15\fm5. The RRc assumption gives a mean (B$-$V)$_0$
colour of 0\fm2-0\fm4 (e.g. Lee \& Carney 1999) resulting in
E(B$-$V)$\approx$0\fm7.
On the other hand, there are a few longer period $\delta$ Scuti stars known
with periods between 0.25 and 0.3 days (Rodr\'\i guez \& Breger 2001), that
could be evolved massive $\delta$ Scuti variables. Thus, V5 can be either a
new field RRc star or a high-amplitude, longer period $\delta$ Scuti star.
In any case, neither V5 nor V6 seems to be physical a member of M37.

\section{Summary}

We have presented the first variable star survey of the rich open cluster
and have discovered seven new variable stars, two pulsating and five
eclipsing. Epochs and periods are determined for three W UMa-type systems
and two pulsating stars. The eclipsing binaries are probably cluster
members, while the pulsating stars are background objects lying much further
away
than M37. V5 might be either a high-amplitude $\delta$ Scuti or a field RRc
star, while V6 is classified unambiguously as high-amplitude $\delta$ Scuti
star. More observations are needed to determine periods for V1 and V2, while
it is an interesting question whether the observed light curve shapes of the
contact binaries are stable over a longer period. All of the three W UMa
systems have light curves with different consecutive maxima suggesting
surface inhomogeneities that can be caused by stellar spot activity.

In addition to the variable star survey, a new colour-magnitude diagram has
been constructed which is based on about 2300 stars. Since the cluster lies
very close to the galactic plane, the field star contamination is fairly
strong. We have determined independently the cluster's main physical
parameters by fitting
isochrones taken from Bertelli et al. (1994). The best fits yield 
E(B$-$V)=0\fm29$\pm$0\fm03, m$-$M=11\fm48$\pm$0\fm13 and ${\rm log}~t$=8.65.
In more convenient units, the cluster's age is about $450\cdot10^6$ years
and the distance to M37 is about 1300 pc. These parametes are in good
agreement with those available in the literature.

\begin{acknowledgements}

This project has been supported by FKFP Grant 0010/2001,
OTKA Grant \#T032258,
``Bolyai J\'anos'' Research Scholarship of LLK from the
Hungarian Academy of Sciences and Szeged Observatory
Foudation. The warm hospitality of the staff of the
Konkoly Observatory and their provision of telescope time
is gratefully acknowledged. Thanks are due to the
referee, Dr. S. Frandsen, whose suggestions led to significant
improvement of the paper. The authors also acknowledge careful
reading of the manuscript by C. Lloyd.
The NASA ADS Abstract Service was used to
access data and references. This research has made use of
the SIMBAD database, operated at CDS-Strasbourg, France.

\end{acknowledgements}

\end{document}